\documentclass[onecolumn]{aastex62}

\graphicspath{{./notes/figs/},{./codes/figs_tess/},{./codes/periodogram/},{./figs2/}}
\usepackage[multidot]{grffile}
\usepackage{comment}
\usepackage{amsmath,amssymb}
\usepackage{bm}

\newcommand{\re}{R_{\rm E}}
\newcommand{\tess}{\textit{TESS}}
\newcommand{\kepler}{\textit{Kepler}}

\newcommand{\gaia}{\textit{Gaia}}
\newcommand{\mbh}{M_\bullet}
\newcommand{\mbhi}{M_{\bullet\mathrm{i}}}
\newcommand{\mbhic}{M_{\bullet\mathrm{i},\mathrm{core}}}
\newcommand{\msun}{M_\odot}
\newcommand{\rsun}{R_\odot}
\newcommand{\ms}{M_\star}
\newcommand{\rs}{R_\star}

\newcommand{\yr}{\mathrm{yr}}

\renewcommand{\d}{\mathrm{d}}

\newcommand{\days}{\mathrm{days}}

\newcommand{\slim}{\sigma_{\rm lim}}

\renewcommand{\sl}{\mathrm{self}\mathchar`-\mathrm{lensing}}
\newcommand{\beam}{\mathrm{beaming}}
\newcommand{\ev}{\mathrm{EV}}

\shorttitle{Black Holes from \tess}
\shortauthors{Masuda and Hotokezaka}

\begin{document}


\title{
Prospects of Finding Detached Black Hole--Star Binaries with \tess
}

\correspondingauthor{Kento Masuda}
\email{kmasuda@astro.princeton.edu}

\author{Kento Masuda}
\altaffiliation{NASA Sagan Fellow}
\affil{Department of Astrophysical Sciences, Princeton University,
Princeton, NJ 08544, USA}

\author{Kenta Hotokezaka}
\affil{Department of Astrophysical Sciences, Princeton University,
Princeton, NJ 08544, USA}



\begin{abstract}

We discuss prospects of identifying and characterizing black hole (BH) companions to normal stars on tight but detached orbits, using photometric data from the \textit{Transiting Exoplanet Survey Satellite} (\tess). We focus on the following two periodic signals from the visible stellar component: (i) in-eclipse brightening of the star due to gravitational microlensing by the BH (self-lensing), 
and (ii) a combination of ellipsoidal variations due to tidal distortion of the star and relativistic beaming due to its orbital motion (phase-curve variation). 
We evaluate the detectability of each signal in the light curves of stars in the \tess\ input catalog, based on a pre-launch noise model of \tess\ photometry as well as the actual light curves of spotted stars from the prime \kepler\ mission to gauge the potential impact of stellar activity arising from the tidally spun-up stellar components.
We estimate that the self-lensing and phase-curve signals from BH companions, if exist, will be detectable in the light curves of effectively $\mathcal{O}(10^5)$ and $\mathcal{O}(10^6)$ low-mass stars, respectively,
taking into account orbital inclination dependence of the signals. 
These numbers could be large enough to actually detect signals from BHs: simple population models predict some 10 and 100 detectable BHs among these ``searchable" stars, although the latter may be associated with a comparable number of false-positives due to stellar variabilities and additional vetting with radial velocity measurements would be essential.
Thus the \tess\ data could serve as a resource to study nearby BHs with stellar companions on shorter-period orbits than will potentially be probed with \gaia.
 
\end{abstract}




\section{Introduction}

Because the most massive stars end their lives as black holes (BHs), stellar mass BHs should exist ubiquitously: the stellar mass function suggests that about $10^8$ BHs exist in the Galaxy \citep[e.g.,][]{1994ApJ...423..659B}, and the nearest ones are expected to be within $\mathcal{O}(10)\,\mathrm{pc}$ \citep{1983bhwd.book.....S, 2003ApJ...596..437C}. Nevertheless, only a part of them have been probed via X-ray/radio emissions from interacting binaries with stellar companions (X-ray binaries) or from pulsars, which are presumably observable only in a fraction of the systems' lifetime and/or the parameter space of such binaries.
The nearest known BH in an X-ray binary is about $1\,\mathrm{kpc}$ away \citep{2010ApJ...710.1127C, 2018arXiv180411349G}, and we likely miss many nearby BHs.

A larger population of stellar mass BHs can be probed if we have a means to search for the quiescent systems with stellar companions on wider, \textit{detached} orbits.
They do not only help completing the census of nearby compact objects, but the visible companions allow for reliable measurements of BH mass and kinematics of the system in the Galaxy, which are both direct probes of the mass loss and kick during the supernova (SN) explosion \citep[e.g.,][]{2017hsn..book.1499C}. If they are in tight orbits, we may also learn how the outcomes of binary interactions depend on the systems' property. In this aspect, elemental abundance of the stellar companion also helps to probe the signature of mass exchange. The information will be useful to understand the formation of compact objects and those in close binaries, such as X-ray binaries \citep[e.g.,][]{2006ARA&A..44...49R} and merging BH binaries 
\citep[e.g.,][]{2016PhRvL.116f1102A, 2016PhRvX...6d1015A}.

Detached BH companions of normal stars can be searched using the same techniques to identify unresolved binaries. The spectroscopic (i.e., radial velocity) search has been considered since 1960s \citep[e.g.,][]{1966SvA....10..251G}, and has recently identified massive, yet dark companions to stars both in a cluster \citep{2018MNRAS.475L..15G} and in the field \citep{2018arXiv180602751T}, whose minimum masses imply that they are BHs or massive neutron stars (NSs). The potential of \gaia\ astrometry has also been discussed extensively \citep{2017MNRAS.470.2611M, 2017ApJ...850L..13B, 2018ApJ...861...21Y, 2018MNRAS.481..930Y}, and hundreds or thousands of BHs may be found by the end of its five-year mission. The typical targets will be $\sim10\,M_\odot$ BHs in au-scale binaries.

This paper focuses on all-sky photometry as another means to search for stars with BH/NS companions: we especially consider the potential of the {\it Transiting Exoplanet Survey Satellite} \citep[\tess,][]{2014SPIE.9143E..20R} to identify and characterize such binaries on tight orbits ($\lesssim0.3\,\mathrm{au}$). To achieve its main science goal to find transiting exoplanets around nearby stars, 
\tess\ will provide photometric light curves with sub-precent precision and at least 27-day long, for $>10^7$ stars in the almost entire sky \citep{2015ApJ...809...77S}.
The number could be sufficiently large to find such rare binaries with compact objects, as shown below. We discuss two methods that have successfully identified white dwarf (WD) companions to normal stars in the photometric data from the \kepler\ mission \citep{2009Sci...325..709B}: (i) periodic brightening due to in-eclipse microlensing known as ``self-lensing" \citep{2014Sci...344..275K, 2018AJ....155..144K, 2019arXiv190707656M}, and (ii) phase-curve modulation due to ellipsoidal variations and Doppler beaming \citep[e.g.,][]{2007ApJ...670.1326Z, 2010A&A...521L..59M, 2015ApJ...815...26F}. 

Those BH--star binaries, if detected with \tess, would necessarily be nearby systems amenable to various follow-up observations, 
with their short-period and repeating signals being ideal for detailed characterization. In particular, the self-lensing BHs, if detected, provide unambiguous evidence for their compact nature, which is otherwise difficult to confirm.
They are also complementary to the BHs detectable with \gaia: they have shorter orbital periods and will be a more sensitive probe of the post-interaction systems that likely followed similar formation paths to X-ray binaries or merging BHs. 

In the following, we estimate how many \tess\ stars will be searchable for BH/NS companions with given masses and orbital periods (Section \ref{sec:searchable}). This information will then be combined with simple population models of BH--star binaries to estimate the number of detectable BHs around those searchable stars (Section \ref{sec:yields}). The estimated yields are checked against the known population of X-ray binaries in Section \ref{sec:xb}. 
We summarize and discuss future prospects in Section \ref{sec:summary}. 

\section{The Expected Signals}\label{sec:signals}

The photometric signal from detached BH--star binaries consists of the following (up to) three periodic components: 
\begin{enumerate}
\item ellipsoidal variations (EVs): tidal force due to BH's gravity changes the geometric shape of the star as well as brightness distribution on the stellar surface,
\item Doppler beaming: relativistic aberration of light, time dilation, and Doppler shift caused by the star's orbital motion produces the light variation proportional to its radial velocity toward the observer, and
\item self-lensing: when the BH eclipses the star, the BH acts as a lens to gravitationally magnify the star.
\end{enumerate}
The former two produce variability in phase with the orbital motion \citep[``phase-curve" modulation; see, e.g.,][]{2017PASP..129g2001S}, while the self-lensing causes pulse-like brightening only during the eclipse. Thus the self-lensing signal has a timescale shorter than the phase-curve variation by a factor of $\rs/a$, where $\rs$ is the star's radius and $a$ is the orbital semi-major axis, and their detectabilities can be discussed separately. 

The amplitudes of all three signals are determined once the binary period $P$, BH mass $\mbh$, star's mass $\ms$ and radius $\rs$, and orbital inclination $i$ are specified (see Section \ref{ssec:signals_formula} for quantitative details). Thus the periodicity of the detected signal, combined with the prior knowledge about the star's mass and radius, gives a handle on the companion's mass to select candidate BHs and NSs. For the phase-curve signals, this practically reduces to searching for companions more massive than the visible ``primary" star: the absence of light from the apparently more massive companion suggests its compactness, which will need to be confirmed with follow-up spectroscopy. For the self-lensing signal, on the other hand, the compactness of the companion is indisputable.\footnote{Even the null detection of spectral features of a massive companion can sometimes be ambiguous. The stellar luminosity is not always a monotonic function of mass, and/or the kinematically detected companion may be a close pair of two stars whose total luminosity is smaller than expected for a single star with the same total mass.} Especially, if the pulses with periods $\lesssim10\,\days$ are detected, the companion is most likely a BH or NS, because WDs on short-period orbits do not usually exhibit self-lensing pulses because of their larger physical radii (e.g., \citealt{2003ApJ...584.1042S}, figure 8 of \citealt{2018AJ....155..144K}). The self-lensing systems are also the best targets to measure BH masses, because of the known orbital inclination and clear physical relation between the pulse height and companion mass (see Section \ref{ssec:signals_formula} below).

Figure \ref{fig:amp_period} plots the amplitudes of each component against the orbital period, for a stellar companion with $\ms=1\,\msun$ and $\rs=1\,\rsun$. The EV signal (proportional to $a^{-3}$) is strongest at $P\lesssim1\,\mathrm{day}$, while at longer periods the beaming or self-lensing signal dominates. These features are shown in the model light curves for the edge-on case in Figure \ref{fig:lc_period}. When the orbit is not nearly edge-on, the central pulse due to self-lensing vanishes.

\begin{figure*}
	\epsscale{0.85}
	\plotone{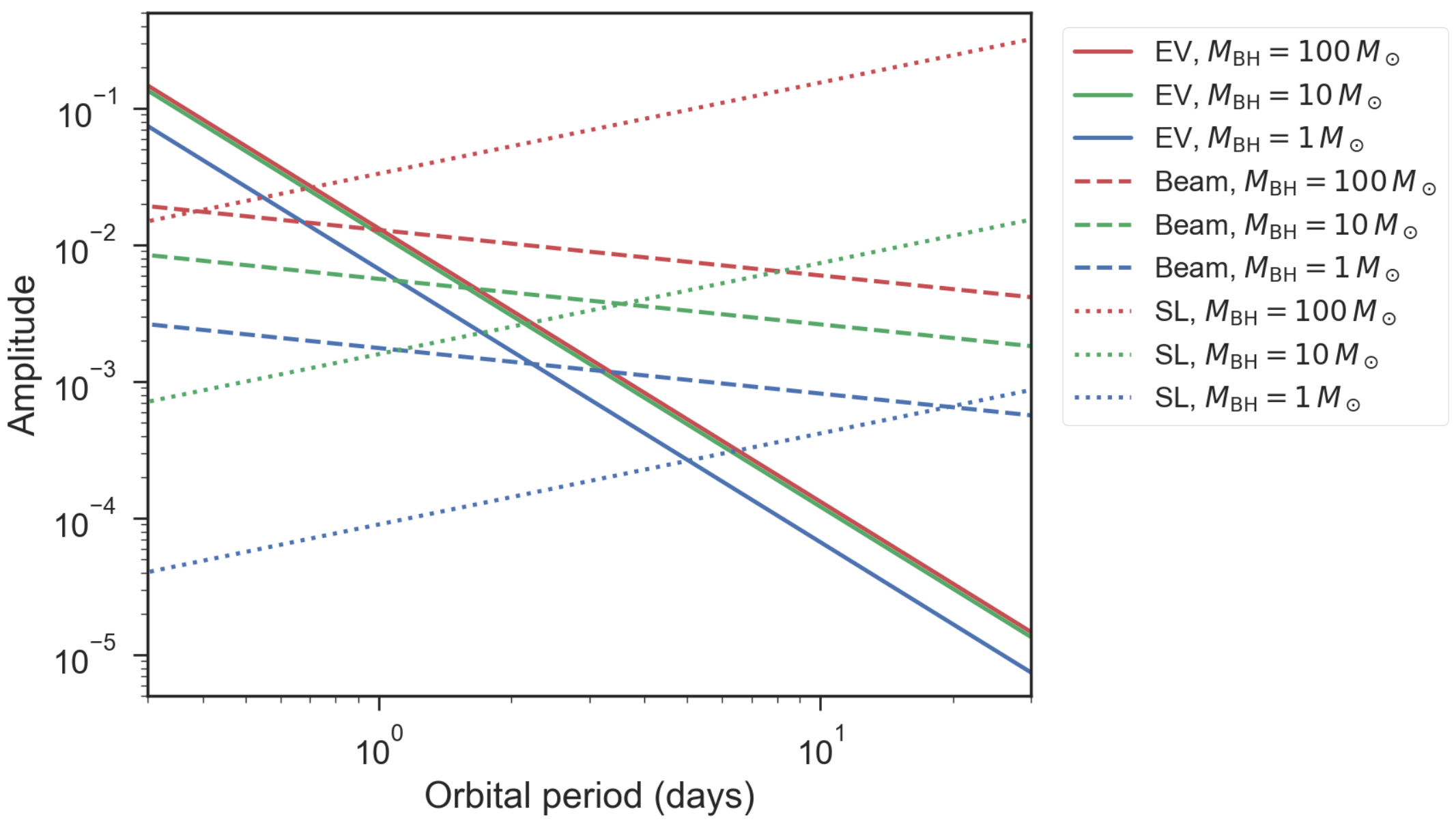}
	\caption{Amplitudes of the EV, beaming, and self-lensing signals versus orbital period for different BH masses. The stellar companion is assumed to be a Sun-like star. \label{fig:amp_period}
	}
\end{figure*}

\begin{figure*}
	\epsscale{0.545}
	\plotone{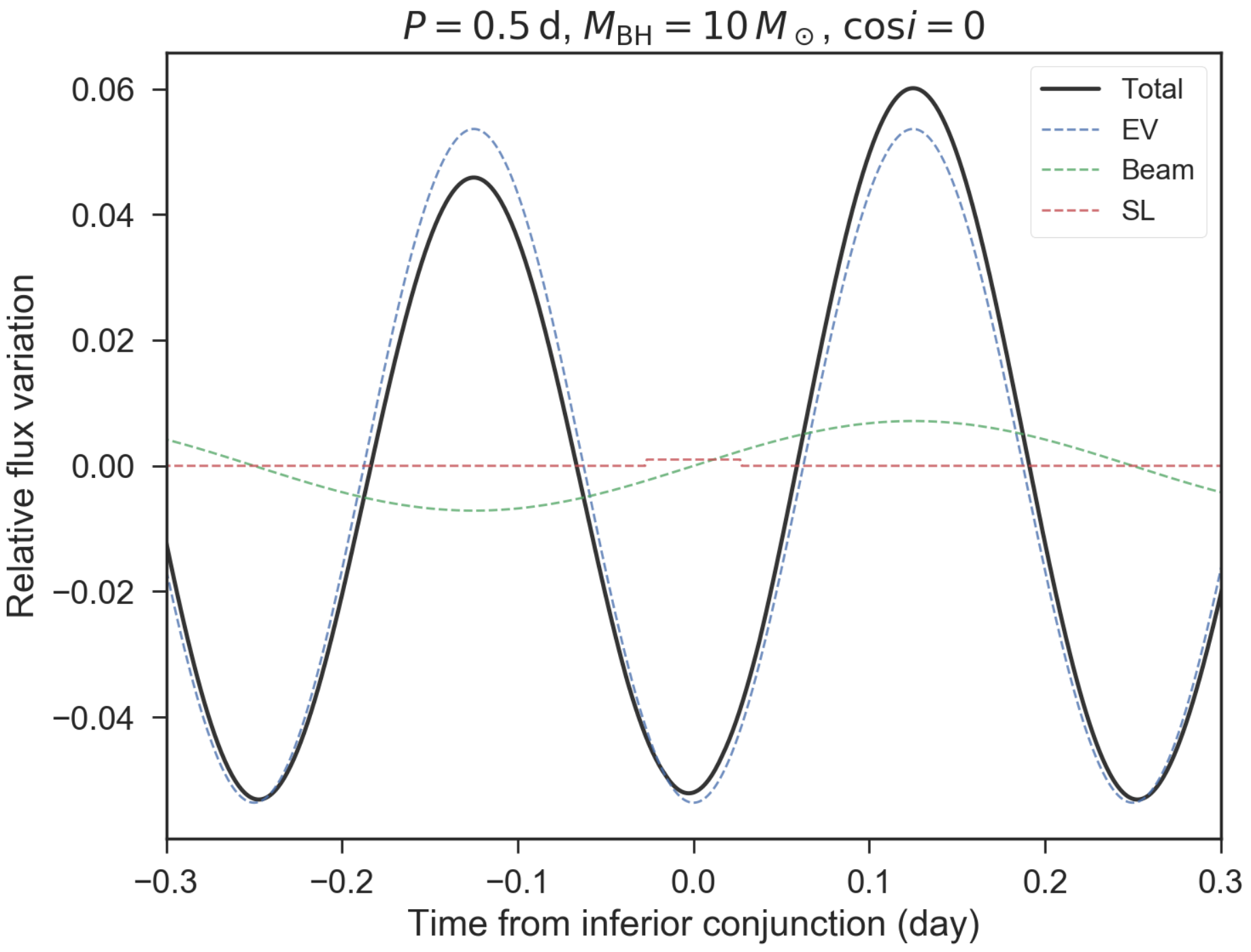}
	\plotone{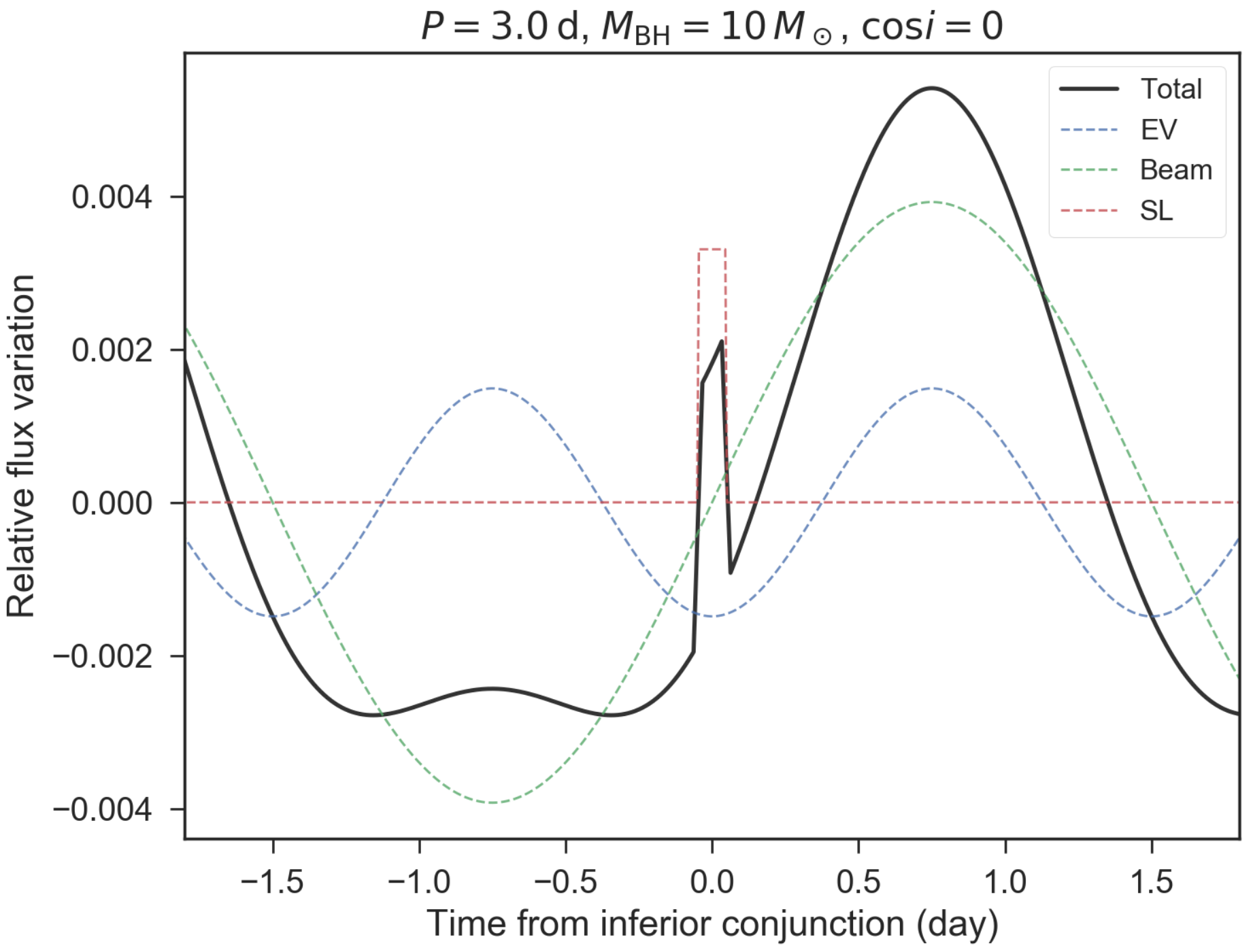}
	\plotone{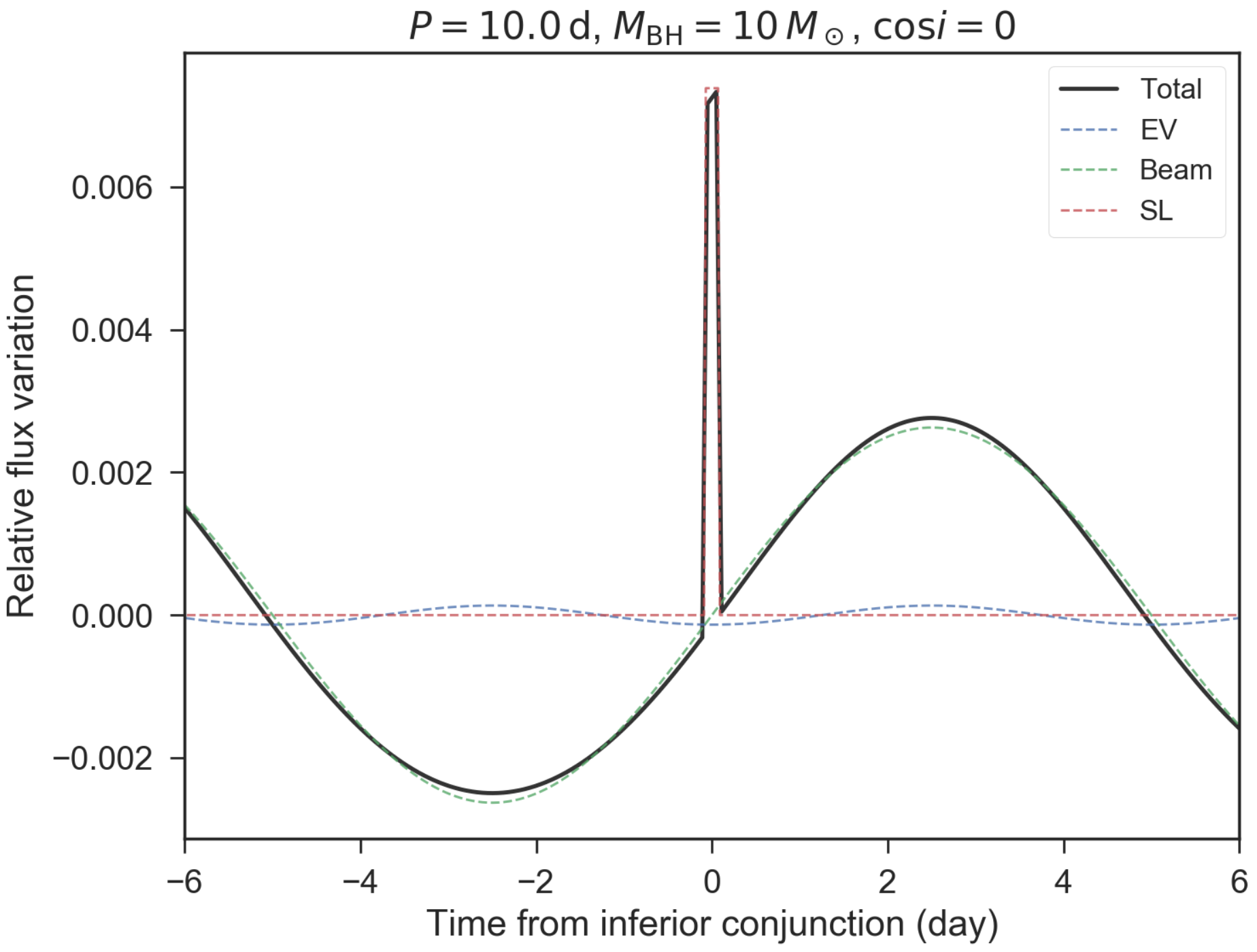}
	\caption{Model light curves ($30$-minute cadence) of detached BH--star systems with different orbital periods: $0.5\,\days$ (\textit{top left}), $3\,\days$ (\textit{top right}), and $10\,\days$ (\textit{bottom}). The primary is a Sun-like star, the BH mass is $10\,\msun$, and the orbit is assumed to be circular and edge-on. When the orbital inclination $i$ is far from $\pi/2$, the ``self-lensing" pulse at time $0$ vanishes.
	\label{fig:lc_period}
	}
\end{figure*}


When only the phase-curve variation is detected, a measurement of the spectroscopic orbit is most likely required to exclude other astrophysical sources of variability including pulsations or star spots. 
While the visible star's location on the Hertzsprung--Russell diagram as available from the \gaia\ data \citep{2018arXiv180409382G} will be helpful to check if the star can be such a pulsator that mimics the detected modulation, star-spot modulations are likely more difficult to distinguish from the phase-curve signal. Because spotted stars are much more common than the stars with BH/NS companions, they could cause too many false positives that make the search based on phase-curve signals to be impractical. Even in the actual BH/NS systems, the star-spot modulations can severely interfere with the phase-curve signals: if the stellar rotation is tidally synchronized with the orbit of a close BH/NS companion, the spot modulations have the same periods as the phase-curve signals, and their activities can even be enhanced due to rapid stellar rotation associated with short orbital periods.
We will examine these issues in Section \ref{sec:searchable}, and find that sufficiently strong phase-curve signals could still be distinguished from spot modulations and serve as a reasonable means to select promising candidates, although false positive rates could be as high as $\sim50\%$.
If both the EV and beaming signals are detected, the consistency of the relative phase and amplitude of the signals, as well as their coherence, will also make them easier to distinguish from stellar activities. Although we assume circular binary orbits throughout this paper, BH--star systems on highly eccentric orbits, if exist, are also good targets because the resulting ``heartbeat" light curve \citep{2012ApJ...753...86T} is easier to distinguish from other astrophysical sources of variability. The shape of the light curve may even yield the orbital inclination to break the degeneracy in the spectroscopic mass \citep{1995ApJ...449..294K}. In our population models in Section \ref{sec:yields}, tides do not necessarily circularize the orbit for $P\gtrsim$ a few days, if the orbit is eccentric after the formation of a BH.

\subsection{Quantitative Descriptions of the Signals}\label{ssec:signals_formula}

\begin{description}
\item[Ellipsoidal variation] Tidal force due to BH's gravity changes the geometric shape of the star as well as brightness distribution on the stellar surface via gravity darkening \citep{1924MNRAS..84..665V}. This causes the light modulation with an amplitude of
\begin{equation}
	\label{eq:ev}
	s_{\rm ev}=\alpha_{\rm ev}\,{\mbh \sin i \over \ms} \left(\rs \over a\right)^3 \sin i
	=1.89\times10^{-2}\alpha_{\rm ev}\sin^2i\,\left({P \over 1\,\mathrm{day}}\right)^{-2}
	\left({\rho_\star \over 1\,\mathrm{g\,cm^{-3}}}\right)^{-1}\left(1\over 1+\ms/\mbh\right).
\end{equation}
Here  
\begin{equation}
	\alpha_{\rm ev}=0.15\,{(15+u)(1+g) \over 3-u},
\end{equation}
where $g$ is the gravity-darkening coefficient and $u$ is the linear limb-darkening coefficient \citep{1993ApJ...419..344M}.
For the circular orbit, the modulation is symmetric with respect to our line of sight, and the signal has the period half the orbital one. In this paper we neglect the dependence on the eccentricity and set $\alpha_{\rm ev}=1$ just for simplicity, since our purpose is not to give a precise prediction for the expected yield.

Figure \ref{fig:amp_period} shows that, at a fixed orbital period, the amplitude of the EV signal saturates with increasing companion masses. This is because $a^3 \sim \mbh$ when $\mbh\gg\ms$ and this dependence cancels the companion mass dependence of the tidal force (Eqn. \ref{eq:ev}). Hence it will be difficult to precisely weigh the most massive companions with the EV amplitude alone.


\item[Doppler beaming] Relativistic aberration of light, time dilation, and Doppler shift caused by the primary's radial acceleration produces the photometric variation with the amplitude of \citep{2003ApJ...588L.117L}
\begin{equation}
	\label{eq:beam}
	s_{\rm beam}=\alpha_{\rm beam}\,4\,{K_\star \over c}
	=2.8\times10^{-3}\alpha_{\rm beam}\sin i\left({P \over 1\,\mathrm{day}}\right)^{-1/3}
	\left( \mbh+\ms \over \msun \right)^{-2/3}\left(\mbh \over \msun\right),
\end{equation}
where $\alpha_{\rm beam}$ is given by integrating the following wavelength-dependent factor
\begin{equation}
	\alpha_{\rm beam,\nu} ={1\over4}\left(3-{{\mathrm{d}\log F_\nu}\over{\mathrm{d}\log \nu}}\right)
\end{equation}
over the photometric bandpass, and $F_{\nu}$ is the spectrum of the star. Again we set $\alpha_{\rm beam}=1$, which is reasonable for a Sun-like star \citep[e.g.,][]{2017PASP..129g2001S}.
This signal is anti-phased with the radial velocity variation of the star and has a different orbital-phase dependence from the ellipsoidal variation (see Figure \ref{fig:lc_period}).
Thus, if both the EV and beaming signals are detected, the consistency of their phases and amplitudes will be useful to confirm the binary origin. An independent measurement of the spectroscopic orbit also helps the interpretation of the light curve because the shape and amplitude of the beaming component are essentially fixed.

\item[Self lensing] If the system is eclipsing, the BH passing in front of the stellar disk acts as a lens to gravitationally magnify the background star \citep[e.g.,][]{1995ApJ...446..541G, 2003ApJ...584.1042S, 2011MNRAS.410..912R}. The light curve exhibits periodic pulses with the height \citep{1969ApJ...156.1013T, 1973A&A....26..215M, 2003ApJ...594..449A}:
\begin{equation}
	\label{eq:sl}
	s_{\rm sl}=2\left(\re \over \rs\right)^2
	=7.15\times10^{-5}\left(\rs \over \rsun\right)^{-2}
	\left({P \over 1\,\mathrm{day}}\right)^{2/3}\left({\mbh \over \msun}\right)\left({\mbh+\ms \over \msun}\right)^{1/3},
\end{equation}
where
\begin{equation}
	\re=\sqrt{{4G\mbh a \over c^2}}
	=4.27\times10^{-2}\rsun \left({P \over 1\,\yr}\right)^{1/3}\left({\mbh \over \msun}\right)^{1/2}\left({\mbh+\ms \over \msun}\right)^{1/6}
\end{equation}
is the Einstein radius. The duration of the signal is
\begin{equation}
	\label{eq:duration}
	\tau_{\rm sl}={\rs P \over \pi a}\cdot {\pi \over 4}
	=1.8\,\mathrm{hr}\times {\pi \over 4}\left(P\over1\,\mathrm{day}\right)^{1/3}\left({\mbh+\ms}\over\msun\right)^{-1/3}\left(\rs\over\rsun\right)
\end{equation}
when averaged over the impact parameter, and the signal vanishes if $\cos i>\rs/a$ for a circular orbit. The pulses have the same period as the orbital period.

Since stellar light is also blocked by the eclipsing compact object, the pulses are not observed unless $\sqrt{2}R_{\rm E}$, which increases with the binary separation, is larger than the physical size of the object. This is why all the known self-lensing binaries containing WDs \citep{2014Sci...344..275K, 2018AJ....155..144K, 2019arXiv190707656M} have orbital periods ranging from months to years, longer than typical eclipsing systems. On the other hand, the physical radius has negligible effects for BHs and NSs, and so they always show pulses regardless of the orbital period. Depending on the detected period the WD case can be excluded by sheer presence of pulses.

If the orbital period is measured from multiple pulses, the BH mass is derived from the pulse height and the stellar radius. For the latter, the parallax information from \gaia\ has already allowed for $10\%$-level measurements for tens of millions of stars \citep{2018A&A...616A...8A}. If combined with the spectroscopic effective temperature, the precision better than five percent can be achieved \citep{2018ApJ...866...99B}; this translates into $10\%$-precision for the BH mass, provided that the pulse height is sufficiently well constrained.
\end{description}

\section{The Number of Searchable Stars}\label{sec:searchable}

Here we estimate how many stars that will be observed by \tess\ 
will enable the detection of the above photometric signals due to BH companions, if exist. 
We focus on the cases where (i) the self-lensing pulse is detectable, and (ii) either the EV or beaming signals is detectable. 
We choose the detection threshold so that the signal can be recovered in the presence of realistic (i.e., both correlated and uncorrelated) noise, and that the expected false positive (FP) rate is reasonably low, as will be discussed in the following subsections. In particular, we focus on the effect of enhanced stellar noise associated with rapid stellar rotation: given the tight orbits and large BH/NS masses of our interest, the rotation of our target stars can well be synchronized with the binary orbit, and stars with such short rotation periods $\lesssim10\,\days$ are known to exhibit enhanced activity. We evaluate this effect using actual light curves of $\approx34,000$ spotted stars from the prime \kepler\ mission \citep{2014ApJS..211...24M}.
We also exclude the systems where the star fills its Roche lobe (to focus on non-interacting systems), as well as the systems where gravitational wave (GW) emission causes orbital decay on a timescale significantly shorter than the star's lifetime.

\begin{figure*}[ht!]
	\epsscale{1.1}
	\plottwo{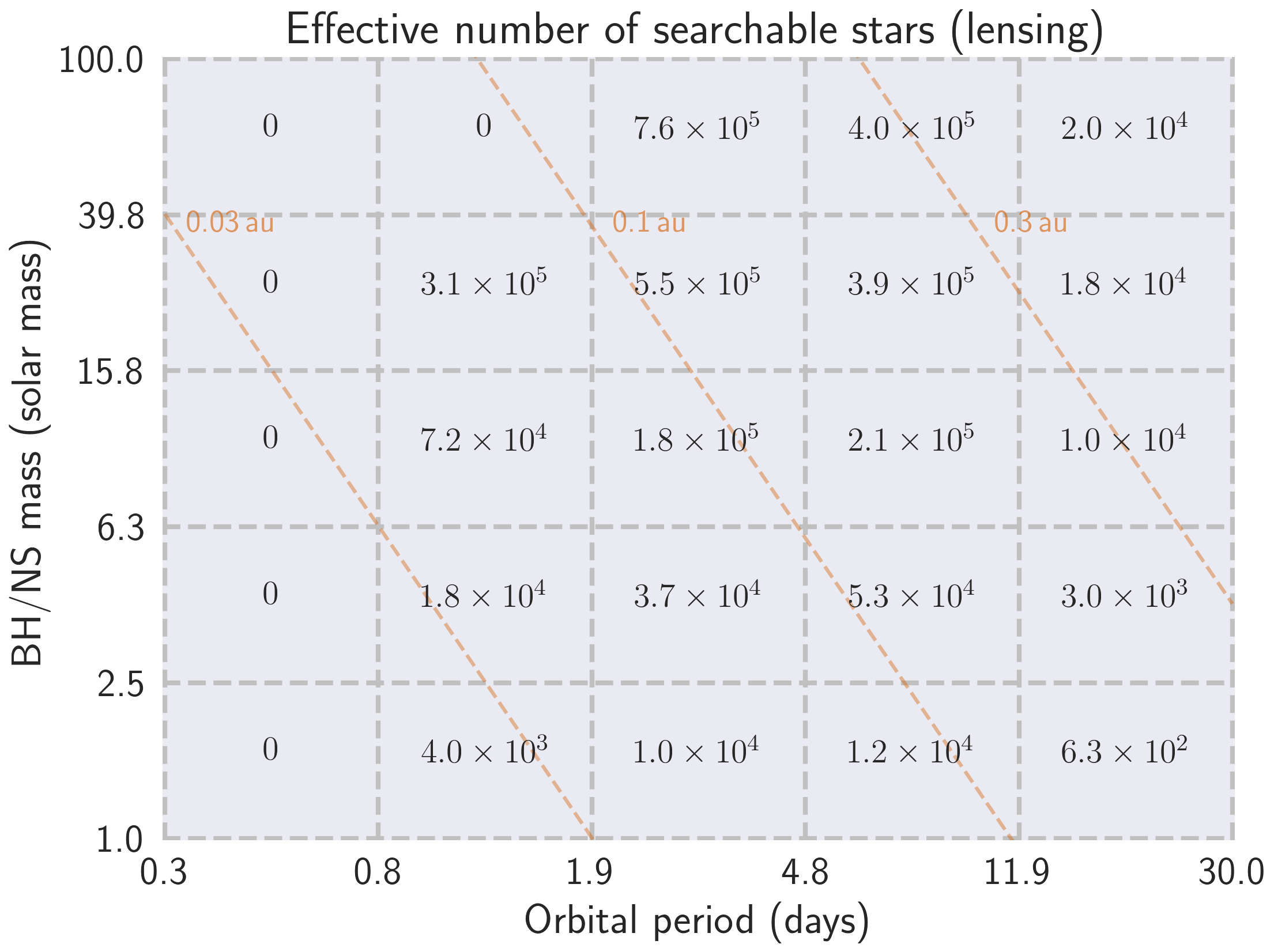}{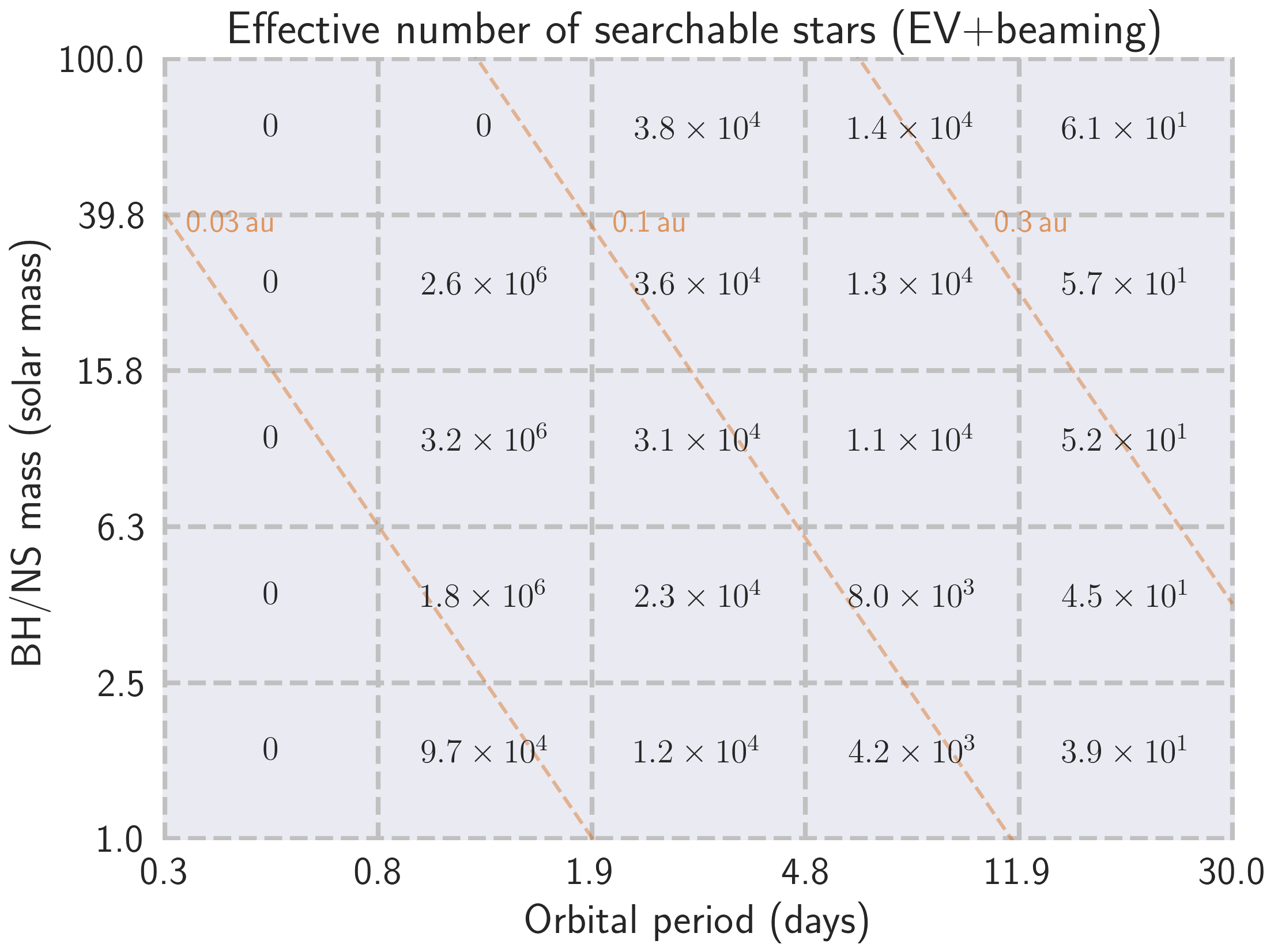}
	\caption{Number of stars in the \tess\ input catalog for which BH companions with given masses and periods are searchable with self-lensing ({\it left}) and phase-curve modulations ({\it right}). Here we count effective numbers of stars, considering the fraction of systems with suitable orbital inclinations assuming that the orbits are isotropic. Here the companion mass is extended down to $1\msun$ to check the searchability of NSs as well. \label{fig:ns} 
	}
\end{figure*}

The number of searchable stars counted this way are shown in Figure \ref{fig:ns} for $5\times5$ log-uniform bins covering the BH/NS mass $1$--$100\,\msun$ and period $0.3$--$30\,\days$; the details of the counting are described below. The numbers are computed for the values at the bin centers, and slant dashed lines show the corresponding values of the semi-major axis assuming that the total mass is dominated by the BH. These numbers take into account the fraction of systems with suitable orbital inclinations: the self-lensing signal can be observed only if the binary orbit is nearly edge-on, and the phase-curve signals gradually become weaker as the system becomes closer to face-on (Eqns.~\ref{eq:ev} and \ref{eq:beam}). This is partly why fewer stars are searchable via self-lensing. We find that self-lensing and phase-curve signals can be used to search effectively $\sim10^5$ and $\sim10^6$ stars, respectively, for stellar companions with periods up to $\sim 10\,\days$. While the EV and beaming signals are mostly sensitive to the shortest-period companions, the detectability of the self-lensing signal has a weaker dependence on the orbital period, because the signal becomes stronger with increasing orbital periods (Eqn.~\ref{eq:sl}).

\subsection{The Stellar Sample and Noise Property}\label{ssec:white_noise}

The main purpose of the \tess\ mission is to find transiting planets around nearby stars. \tess\ employs four cameras each with a field of view (FOV) of $24^{\circ}\times24^{\circ}$, which point each sector of the sky for $27.4\,\days$ (two spacecraft orbits). Over its two-year mission, 26 sectors with partial overlaps will be observed to tile the almost entire sky excluding the region near the ecliptic; the region near the ecliptic poles will be most densely covered and continuously observed for one year. \tess\ will provide images stacked at two-minute cadence for a few $10^5$ targets pre-selected for planet search and asteroseismology, as well as 30-minute cadence full-frame images for all sources.

To enable the selection of optimal targets for the planet search, a catalog of luminous sources on the sky, \tess\ Input Catalog \citep[TIC,][]{2017arXiv170600495S}, has been created. 
We adopt the magnitude in the \tess\ bandpass, stellar mass $\ms$, stellar radius $\rs$, and stellar effective temperature $T_\star$ in the TIC version 6 \citep{2017arXiv170600495S}. We focus on the objects classified as stars, on ecliptic latitude $|l|>6^\circ$ (i.e., outside the gap in the \tess\ FOV), and with both mass and radius estimated in the catalog. The last selection limits the targets from $\sim400$ million to $\sim20$ million low-mass stars (mostly $<2.5\,\msun$) with \tess\ magnitudes brighter than $15$. We do not consider this to be a serious limitation because the quality of the \tess\ photometry will be limited for fainter stars in any case.

We use the fitting formula in \citet[][ver.~20170531]{2017arXiv170600495S} 
to evaluate the white component of
the expected photometric noise as a function of the \tess\ magnitude presented in \citet{2015ApJ...809...77S}. 
We adopt the observing duration $T=27.4\,\days$ (i.e., minimum observing duration) for all the stars to give a conservative estimate. 
We always assume $30\mathchar`-\mathrm{min}$ cadence observations to deal with all the stars that will be in the full-frame images.

\subsection{Searchability for Self-lensing Signals}\label{ssec:searchable_sl}

If the stellar rotation is synchronized with the orbit, the star may show enhanced spot activity with a similar period to that of the self-lensing pulses. Since the duration of self-lensing pulses is much shorter than the orbital period (and hence rotation period of the spin-synchronized star), the rotational modulation can basically be filtered out without affecting the self-lensing signal. However, this removal may not be perfect, and also stellar variabilities may enhance correlated noise on a timescale similar to the duration of self-lensing events. Thus, estimates considering the uncorrelated noise alone (as described in Section \ref{ssec:white_noise}) most likely overestimate the detectability of the self-lensing signal.

To evaluate the detectability in the presence of such effects, we perform injection-and-recovery simulations using the actual light curves of spotted stars from the prime \kepler\ mission. Because \kepler\ has a $\sim$10x larger aperture than \tess, the \kepler\ light curves include much smaller photon noise and can be considered as a realistic template of astrophysical variability of the stars. Therefore we can simulate the light curves of spotted stars observed with \tess\ by adding white noise component due to its smaller aperture (as described in Section \ref{ssec:white_noise}) to the actual \kepler\ light curves.\footnote{This may overestimate the noise for the brightest stars observed by \tess, but those stars constitute only a small fraction and would not affect the detectability of such rare systems as BH/NS binaries.} Here we assume that the stellar rotation is always synchronized with the orbit of the self-lensing systems, and 
inject self-lensing signals with the same period as the rotation period of each \kepler\ star.

We take random segments of $27.4\,\mathrm{day}$ long from the \kepler\ light curves of rotating stars and detrend them with median filters with the length of $0.25\times(P/30)^{1/3}\,\days$, which is typically a few times longer than self-lensing durations and much shorter than rotation period. Then we run the box least-squares (BLS) algorithm\footnote{We used the python implementation {\tt bls.py} by Daniel Foreman-Mackey: {\url https://github.com/dfm/bls.py}.} \citep{2002A&A...391..369K} on these light curves, after randomly injecting Gaussian noise with log-uniform dispersions spanning from 100 to 10,000 ppm (to cover the expected noise range from \tess) and flipping the light curves around their median values (to evaluate any possible source of positive excursion). We then set the threshold value for the peak signal-to-noise of the BLS spectrum, so that no detection is found from $10^6$ trials to achieve false-positive rate less than $10^{-6}$. Given that $\sim10\%$ Sun-like stars exhibit spot modulation detectable with \kepler\  \citep{2014ApJS..211...24M}, this yields the FP rate of $\sim10^{-7}$, which corresponds to $\mathcal{O}(1)$ FP in our sample of $20$M stars. Considering the strong dependence of the spot activity on the rotation period, we choose different thresholds for each period bin, and the resulting thresholds turn out to be higher for shorter-period systems. We also count the BLS peaks as detections only if the detected period is close to the measured rotation period of the stars.\footnote{Here we implicitly assume that the rotation period of the star showing spot modulations can be measured well: this assumption may not hold for longer-period binaries in the 27.4-day data, but the detectability turns out to be low for such systems in any case.}
Then we perform the simulations injecting both Gaussian noise and self-lensing pulses of various amplitudes (100--$10^5$ppm; see Figure \ref{fig:amp_period}) and phases, and check what fraction of the injected signals are recovered using the thresholds as defined above. We fit the resulting recovery rate as a function of the signal-to-noise of the phase-folded signal $\sqrt{T/P}(s_{\rm sl}/\sigma_{\rm 30min})$ using the gamma cumulative distribution function \citep{2017AJ....154..109F} scaled by the asymptotic value of the recovery rate at high signal-to-noise ratios: the values were $\approx100\%$ for $P=0.8$--$4.8\,\days$, $\approx90\%$ for $P=0.3$--$0.8\,\days$ and $P=4.8$--$11.9\,\days$, and $\approx20\%$ for $P=11.9$--$30\,\days$. The resulting function $C_{\rm rec}(\sqrt{T/P}\,(s_{\rm sl}/\sigma_{\rm 30min}))$ is used to evaluate the expected number of searchable stars.

For a star with mass $\ms^j$ and radius $\rs^j$, the effective searchability for a BH/NS companion with period $P$ and mass $\mbh$ on randomly oriented orbits is evaluated as:
\begin{equation}
	\label{eq:neff_sl}
	N_{\rm eff}(P, \mbh, \ms^j, \rs^j)={\rs \over a}\,C_{\rm rec}\left(\sqrt{n}\,{s_{\rm sl}\over \sigma_{\rm 30min}}\right),
\end{equation}
where $\rs/a$ corresponds to the eclipse probability for a circular orbit, and $n$ is the number of pulses in the data. The number $n$ can be either $n_0\equiv[T/P]$ or $n_0+1=[T/P]+1$ depending on the orbital phase, where $[x]$ is the greatest integer that does not exceed $x$. The corresponding probabilities are $p_{n_0}=[T/P]+1-T/P$ and $p_{n_0+1}=T/P-[T/P]$, respectively. We count the number of searchable stars for both $n_0$ and $n_0+1$, and average them with the weights $p_{n_0}$ and $p_{n_0+1}$. We set $p_{n_0=1}=0$ to count only the pulses that are observed at least twice, although such longer-period systems turned out to be difficult to search in any case.

\subsection{Searchability for Phase-curve Signals}

The interference with stellar activity is a more serious issue for the phase-curve signal, because the star-spot modulation includes signals with both periods $P/2$ and $P$, if the rotation is synchronized with the orbit. This means that simple periodicity search will introduce a large number of FPs from spot modulations, because there exist more spotted stars than the BH/NS systems. Here we consider a search focusing on EV/beaming signals that are likely too large to be produced by a spotted star with a given rotation period and mass, and examine if the search could provide reasonable number of candidates without being completely swamped by false positives.

To set the threshold, we compute Fourier transform of the \kepler\ light curves of all the spotted stars in \citet{2014ApJS..211...24M} and measure the amplitudes of the peaks corresponding to the rotation period $s_P$ and its first harmonic $s_{P/2}$. We then divide the sample into the period bins as we used for orbital periods, as well as temperature bins ($<4000\,\mathrm{K}$, 4000--5000$\,\mathrm{K}$, 5000--6000$\,\mathrm{K}$, $>6000\,\mathrm{K}$), and take the maximum values of $s_P$ and $s_{P/2}$ in each bin. We set the threshold for the EV and beaming signals, $s_{\rm ev,th}$ and $s_{\rm beam,th}$, so that the Fourier amplitudes from the EV and beaming signals are larger than $s_{P/2}$ and $s_P$, respectively. This allows us to pick up phase-curve signals that are likely too strong to be caused by spot modulations, taking into account the dependence of their amplitudes on the stellar rotation period and effective temperature.
These thresholds are much larger than the noise level and statistical false positives are negligible. Thus, assuming that Sun-like stars observed by \tess\ have similar properties as the \kepler\ sample, the FP rate due to spotted stars would roughly be less than one in $10^5$. The rate corresponds to some $100$ FPs in our 20M sample stars.\footnote{
Since \kepler\ has observed only $10^5$ stars, we cannot estimate the threshold that guarantees a lower FP rate with this method.
} 
Although this number of possible FPs is not small, if a similar (or larger) number of BH/NS candidates can be detected with this choice of thresholds, the search would still provide meaningful information to select promising candidates, especially considering that further vetting (e.g., presence or absence of any RV variations) would readily be possible for many stars using the archival spectroscopic data \citep[e.g.,][]{2018arXiv180602751T, 2019ApJ...872L..20G}. We will see that this can be the case in Section \ref{ssec:yields_results}.

Following Eqns.~\ref{eq:ev} and \ref{eq:beam}, the above detection criteria, $s_{\rm ev}>s_{\rm ev,th}$ and $s_{\rm beam}>s_{\rm beam,th}$ translate into
\begin{equation}
	1-\cos^2 i > {s_{\rm ev,th} \over s_{\mathrm{ev},\cos i=0}} \equiv \beta_\mathrm{ev},
	\quad
	\sqrt{1-\cos^2 i } > {s_{\rm beam,th} \over s_{\mathrm{beam},\cos i=0}} \equiv \beta_\mathrm{beam},
\end{equation}
where the subscript ${\cos i=0}$ refers to the amplitude for a system with $\cos i=0$. 
Thus, averaging over random orbital inclination (i.e., uniform $\cos i$), the effective searchability of each star is given by $H(1-\beta_\mathrm{ev})\sqrt{1-\beta_\mathrm{ev}}$ for the EV signal and $H(1-\beta_\mathrm{beam})\sqrt{1-\beta_\mathrm{beam}^2}$ for the beaming signal, where $H$ is the Heaviside step function ($H(x)=1$ for $x>0$ and $0$ otherwise). Considering that either of the signal, if above the threshold, would be sufficient to flag the system to be anomalous, we take the larger of the two values to evaluate the searchability of each star:
\begin{equation}
	\label{eq:neff_phase}
	N_{\rm eff}(P, \mbh, \ms^j, \rs^j, T_\star^j)=\mathrm{max}\left[ H(1-\beta_{\rm ev})\sqrt{1-\beta_{\rm ev}},\ H(1-\beta_{\rm beam})\sqrt{1-\beta_{\rm beam}^2} \right].
\end{equation}

\subsection{Counting Searchable Stars}

Since we focus on detached systems, we exclude the stars filling their Roche lobes. We compute the effective Roche-lobe radius for the star by \citep{1983ApJ...268..368E} 
\begin{equation}
	\label{eq:roche}
	R_{\rm L}(q,a)={0.49q^{2/3} \over {0.6q^{2/3}+\ln(1+q^{1/3})}}\,a
	\equiv \tilde R_{\rm L}(q)\,a,
\end{equation}
where $q=\ms/\mbh$.  We require that $\rs<R_{\rm L}$, or $a>\rs/\tilde R_{\rm L}$.
We also exclude the systems where the GW emission causes rapid orbital decay compared to the main-sequence lifetime of the star. We compute the decay time as $t_{\rm GW}=3.3\times10^8\,\mathrm{Gyr}\times{(a/\mathrm{AU})^4 \over \mbh\ms(\mbh+\ms)/\msun^3}$ \citep{1964PhRv..136.1224P} 
and the main-sequence lifetime as $t_{\rm MS}=10\,\mathrm{Gyr}(\ms/\msun)^{-2.5}$, and omit the systems with $t_{\rm GW}<t_{\rm MS}/2$.
Taking all these into account, the number of searchable stars for a set of $(P, \mbh)$ is computed by:
\begin{align}
	N_{\rm searchable}(P, \mbh)= \sum_j N_{\rm eff}(P, \mbh, \ms^j, \rs^j, T_\star^j)\cdot H(a^j-\rs^j/\tilde R^j_{\rm L})\cdot H(t_{\rm GW}^j-t_{\rm MS}^j/2),
\end{align}
where the index $j$ runs over all the stars and $N_{\rm eff}$ is computed for each of the self-lensing signal (Eqn.~\ref{eq:neff_sl}) and phase-curve signal (Eqn.~\ref{eq:neff_phase}). The results are shown in Figure \ref{fig:ns} for the self-lensing signal (left) and for the phase-curve signal (right).

\section{Expected Number of Detectable Black Holes}\label{sec:yields}

Here we estimate how many detectable BH companions actually exist around the above searchable stars. To do so we multiply the number of searchable stars by the intrinsic occurrence of BH companions as a function of the BH/stellar mass and binary orbital period. More specifically, given the probability density that a star with mass $\ms$ has a BH companion with period $P$ and mass $\mbh$, $p(P, \mbh|\ms)$, the number of detectable companions per unit $P$ and $\mbh$, $n_{\rm det}(P, \mbh)$, is given by:
\begin{equation}
	n_{\rm det}(P, \mbh)\Delta P\Delta\mbh 
	=\sum_j N_{\rm eff}(P, \mbh, \ms^j, \rs^j, T_\star^j) \cdot p(P, \mbh|\ms^j)
	\Delta P\Delta\mbh,
\end{equation}
where the index $j$ runs over the searchable stars in each $\Delta P\Delta\mbh $ bin.

The occurrence of short-period BH--star binaries $p(P, \mbh|\ms)$ is quite uncertain both theoretically and observationally. So we adopt two simple models to provide crude estimates in the two extreme cases, with and without binary evolution. These estimates are thus not meant to be a precise prediction, but are rather to provide an order-of-magnitude sense of feasibility.

\subsection{Population Models of BH--Star Binaries}\label{ssec:yields_bhpops}

\subsubsection{Estimate Based on Field Binaries}\label{sssec:yields_bhpops_field}

Here we construct the population of BH--star binaries assuming that their properties follow those of field binaries \citep[cf.][]{2017MNRAS.470.2611M}.
We pick up ``BHs"  from a power-law mass function, $I_{\rm BH}(\mbh)\propto \mbh^{-2.3}$, with a minimum mass of $5\,M_{\odot}$.
The mass function is normalized such that all the stars with $\geq 20M_{\odot}$ end up as BHs.
This gives the occurrence of BHs as:
\begin{equation}
	{\d N_{\rm BH} \over \d \mbh}=H(\mbh-5\,\msun)\,I_{\rm BH}(\mbh).
\end{equation}
Then we assign companion stars to these BHs, based on the occurrence of massive star binaries in the field (e.g., \citealt{2012Sci...337..444S}):
\begin{equation}
	f_{\rm companion}(q, P)\,\d q\,\d P= C\,{q^0\over P}\,\d q\,\d P, \label{eq:comp}
\end{equation}
where $q\leq1$ is the binary mass ratio, $P$ is the orbital period, and $C$ is the normalization constant that fixes the binary fraction (see below). Thus the occurrence of BHs with stellar companions is:
\begin{equation}
	{\d N_\mathrm{BH\mathchar`-star} \over {\d\mbh\,\d\ms\,\d P}}
	={1\over\mbh}{\d N_\mathrm{BH\mathchar`-star} \over {\d\mbh\,\d q\,\d P}}
	={1\over\mbh}\,{\d N_{\rm BH} \over \d \mbh}\,f_{\rm companion}\left({\ms\over\mbh}, P\right)
	={C\over\mbh}\,{\d N_{\rm BH} \over \d \mbh}\,{1\over P}.
\end{equation}
Here we assume circular orbits for simplicity. 
This yields
\begin{equation}
	\label{eq:bhocc}
	p_{\rm field}(P, \mbh| \ms)
	={{\d N_\mathrm{BH\mathchar`-star}/{\d\mbh\,\d\ms\,\d P}}\over \d N_{\rm star}/\d\ms}
	={C\over\mbh}\,{H(\mbh-5\,\msun)\,I_{\rm BH}(\mbh) \over I(\ms)}\,{1\over P},
\end{equation}
where $\d N_{\rm star}/\d\ms=I(\ms)$ is the initial stellar mass function from \cite{2001MNRAS.322..231K}.  
We choose $C$ so that the binary fraction integrated over $P=0.1\,\mathrm{days}$ to $P=10^{3.5}\,\mathrm{days}$ is $0.5$, following \citet{2012Sci...337..444S}. Strictly speaking, it is not clear whether the normalization remains the same for binaries containing BHs. The choice is for an optimistic scenario in which formation of a BH does not lead to any loss of binary companions. 

\subsubsection{Estimate Based on a Simple Model of the Common-envelope Evolution}\label{sssec:yields_bhpops_ce}

The above estimate does not take into account any interactions in the binary. If they go through the common-envelope (CE) phase, the low-mass companion may merge with the BH progenitor. Even if it survives, the binary orbit should dramatically shrink during the process. Here we examine the outcome of this CE evolution. Unlike in \citet{2018ApJ...861...21Y}, we do not consider the case where the mass transfer occurs stably, since our focus is on systems with initially large mass ratios.

We first follow the same procedure as in Section \ref{sssec:yields_bhpops_field} to sample the population of binaries consisting of a BH progenitor with mass $\mbhi$ and its stellar companion with mass $\ms$: $p(P_{\rm i}, \mbhi|\ms)$.
Then we use the fitting formulae provided by \cite{2000MNRAS.315..543H} to compute the core mass $\mbhic$ of the BH progenitor as well as its maximum radius $R_{\rm max}$ (typically $\sim 1000$--$3000R_{\odot}$) achieved during its evolution.
We then assume that all the sampled binaries with initial semi-major axes $a_{\rm i}=\left[P_{\rm i}^2G(\mbhi+\ms)/4\pi^2\right]^{2/3}$ smaller than $R_{\rm max}$ goes through the CE phase. The companion survives if the orbital energy is sufficiently large to completely strip the envelope of the BH progenitor; we assume only the stripped core of the BH progenitor is left when this happens. 
The semi-major axis after this process, $a_{\rm f}$, is computed by \citep{1984ApJ...277..355W}:
\begin{equation}
	\alpha \left( {G\mbhic\ms \over 2a_{\rm f}}  - {G\mbhi\ms \over 2a_{\rm i}}\right)
	= {G(\mbhi-\mbhic)\mbhi \over \lambda R_{\rm Roche,i}},
\end{equation}
where $R_{\rm Roche,i}=R_{\rm L}(\mbhi/\ms, a_{\rm i})$ with $R_{\rm L}$ defined in Eq.~\ref{eq:roche} and we adopt $\alpha \lambda = 1$ \citep{2002ApJ...572..407B}.
Finally, we assume that the remnant core of the BH progenitor loses mass via stellar wind and SN explosion before it becomes a BH.
For the wind mass loss during the Wolf--Rayet star phase we follow \citet{2017A&A...607L...8V}. 
At the core collapse, the remnant BH mass and SN ejecta mass are  determined by the formula in \cite{2002ApJ...572..407B}. Both mass loss processes change the semi-major axis.

The above procedures define a transformation between $(P_{\rm i}, \mbhi)$ and $(P, \mbh)$.
The final distribution after the evolution is then computed by:
\begin{equation}
	p_{\rm CE}(P, \mbh|\ms)
	= \left| {\partial (P_{\rm i}, \mbhi) \over \partial(P, \mbh)}\right|\,p_{\rm field}(P_{\rm i}, \mbhi| \ms).
\end{equation}

\subsection{Results}\label{ssec:yields_results}

\begin{figure*}[ht!]
	\epsscale{1.1}
	\plottwo{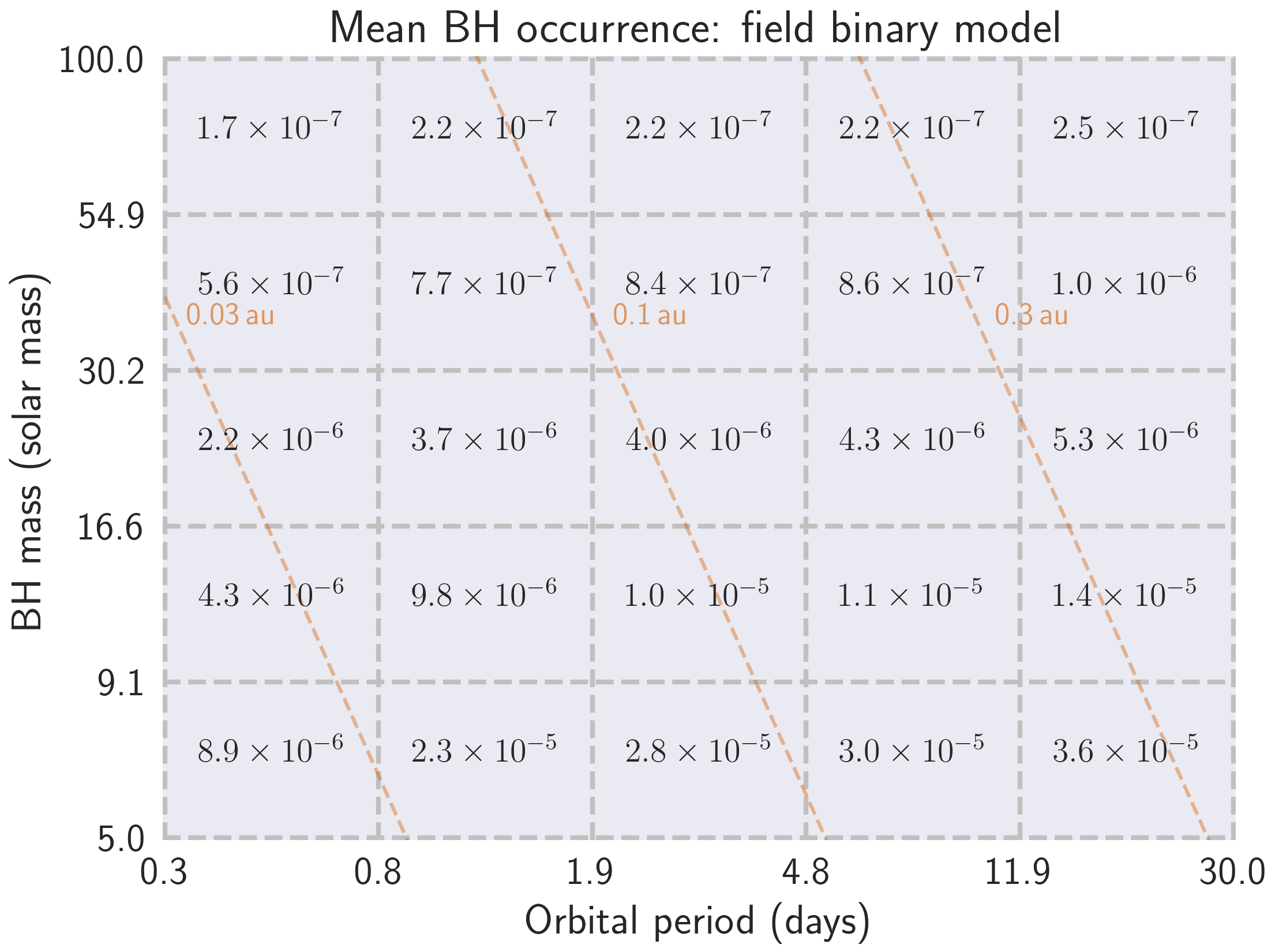}{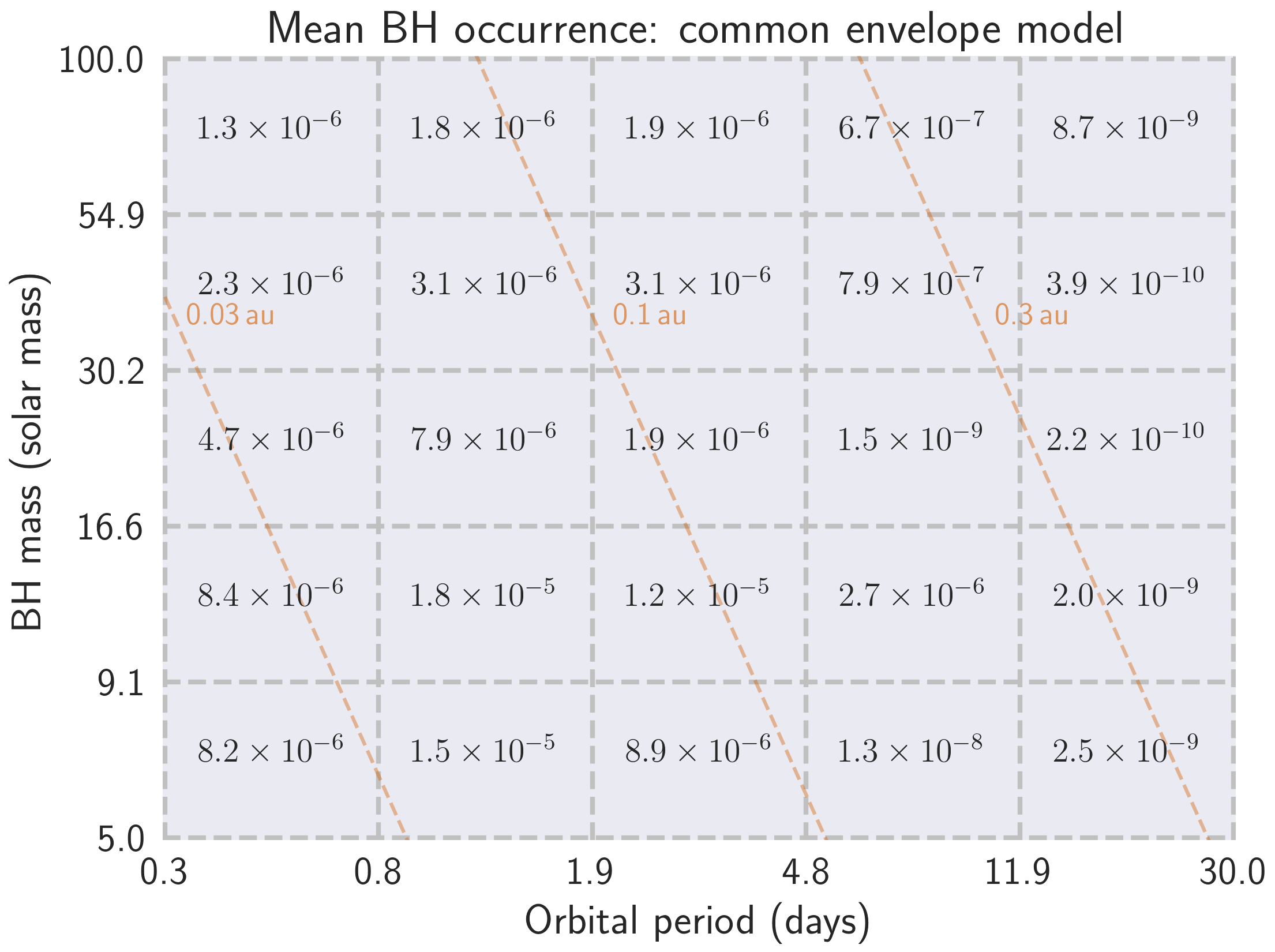}
	\caption{Occurrence rates of BH companions as a function of the BH mass and orbital period from ({\it left}) field-binary model and ({\it right}) CE model with $\alpha\lambda=1$. The values are computed for stellar companions searchable with self-lensing.\label{fig:bhpops}
	}
\end{figure*}

\begin{figure*}[ht!]
	\epsscale{1.1}
	\plottwo{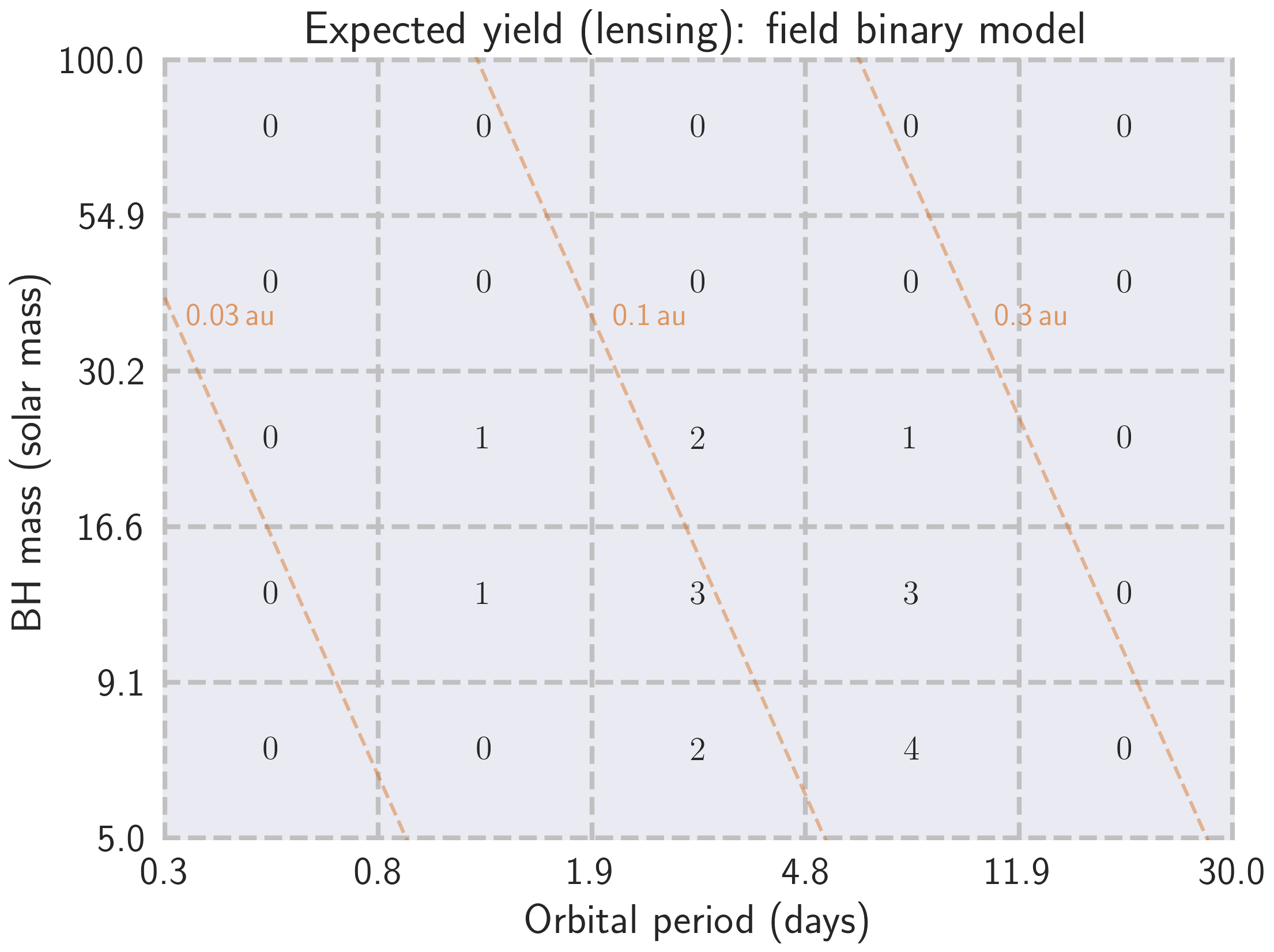}{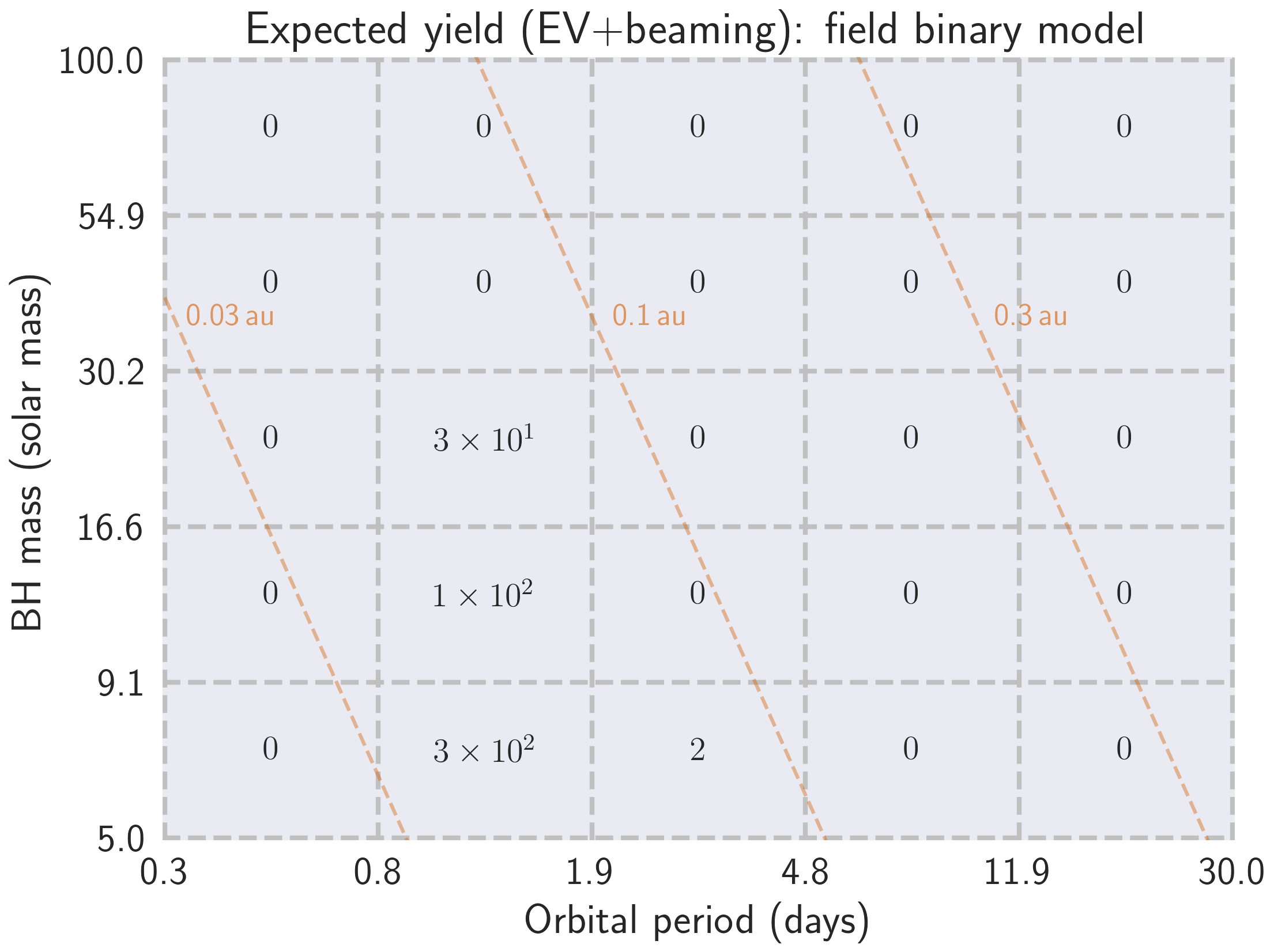}
	\caption{Expected numbers of detectable BH companions for the field binary model, using ({\it left}) self-lensing and ({\it right}) phase-curve modulation. \label{fig:yield}
	}
\end{figure*}

\begin{figure*}[ht!]
	\epsscale{1.1}
    \plottwo{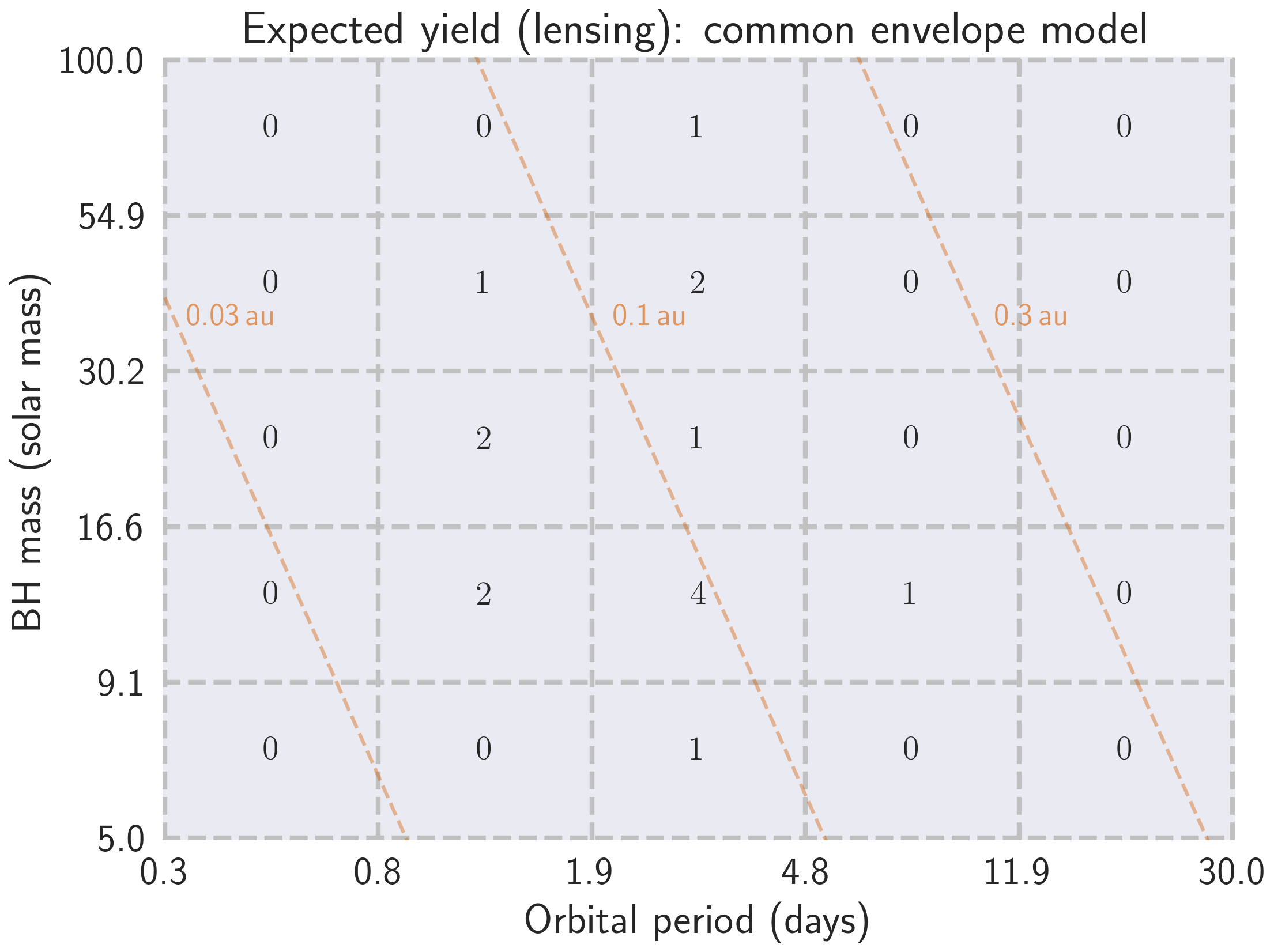}{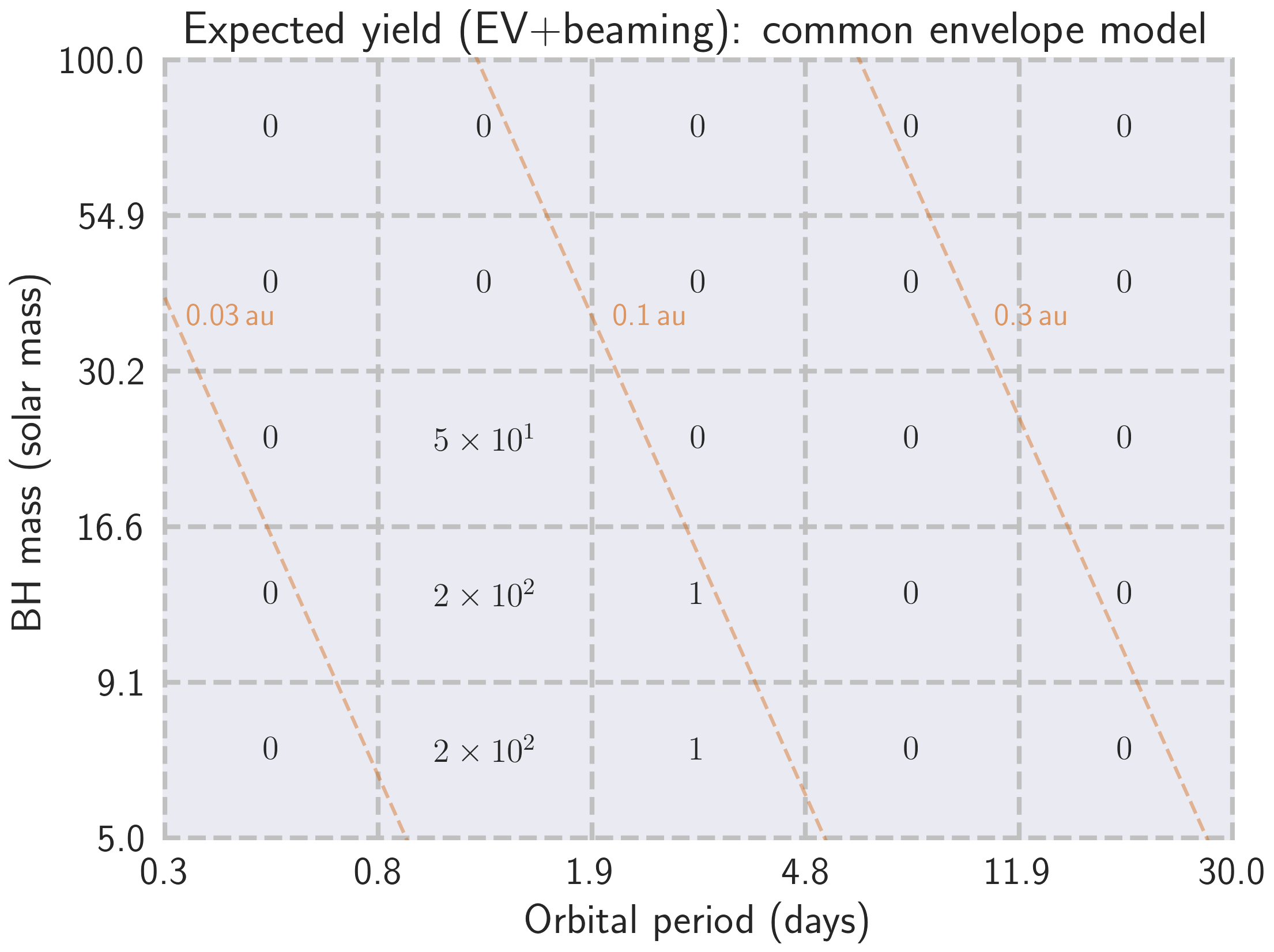}
	\caption{Same as Figure \ref{fig:yield}, but for the CE model with $\alpha\lambda=1$. \label{fig:yield_ce}
	}
\end{figure*}

Figure \ref{fig:bhpops} shows the BH occurrence rates for the field binary model and the CE model with $\alpha\lambda=1$ described in Section \ref{ssec:yields_bhpops}. The values are averages of $p(P, \mbh|\ms)\Delta\mbh\Delta P$ over $\ms$ of the stars searchable with self-lensing in each bin; the differences are mostly within a factor of a few when the stars searchable with phase-curve signals are used instead.
Figures \ref{fig:yield} and \ref{fig:yield_ce} show the estimated numbers of detectable BHs based on the occurrences in Figure \ref{fig:bhpops} and the number of searchable stars in Figure \ref{fig:ns}. 
Both field binary and CE models predict some 10 BH companions detectable  with self-lensing, and some 100 with phase-curve variations. This is the consequence that 
the BH occurrence in the relevant region of the parameter space is about $10^{-4}$ (Figure \ref{fig:bhpops}),
while the effective numbers of stars searchable with self-lensing and phase-curve signals are $\sim10^5$ and $\sim10^6$, respectively. 
Curiously, the estimated occurrence appears to be compatible with the discovery of a BH/NS system by \citet{2018arXiv180602751T} among $\gtrsim10^5$ stars from Apache Point Observatory Galactic Evolution Experiment \citep[APOGEE,][]{2017AJ....154...94M}, although the period we focus on is much shorter.
While the resulting occurrences are similar between the field binary and CE models, the detectable binaries from the CE models were initially in much wider orbits ($P=10^2$ to $10^4\,\days$) and surrendered their orbital energies to survive the CE evolution.

These calculations suggest that the number of searchable stars with \tess\ may indeed be sufficient to actually detect BH/NS companions. This also indicates that even an upper limit on the occurrence rate from the null detection, if quantified, will be able to provide a meaningful observational constraint on the poorly-understood process of the CE evolution.
We also note that the searchability increases both in terms of the number and period range, if individual stars are observed for a longer duration in the overlapping regions of the observing sectors or during a potential extended mission \citep{2017arXiv170508891B}.

In these population models, the typical mass of the stars with detectable BH companions is $1$--$2\,\ms$, mainly by construction of the sample. The brightest star with a BH companion is predicted to have a $V$-band magnitude of $\sim 10$ for the self-lensing population, and will be even brighter for the phase-curve sample. These magnitudes correspond to the distance of a few tens to a few hundreds of parsecs. If actually found, these systems will therefore be among the nearest known BHs.

\subsection{Application to the \kepler\ Sample}

For a sanity check, we applied the same procedure to the \kepler\ sample, using stellar properties and the combined differential photometric precision \citep{2010ApJ...713L..79K} of the stars in the \kepler\ input catalog \citep{2017ApJS..229...30M}. Adopting the same detection thresholds, we found that no detection is expected with either self-lensing or phase-curve signals, consistently with the null detection of such systems so far. The conclusion for the self-lensing BH systems is insensitive to the adopted threshold because the null detection is expected simply due to their low occurrence rate, as pointed out by \citet{2003ApJ...594..449A}. The interpretation of the null detection of phase-curve signals is more subtle. Our population models predict that there would be some 10 such systems on close-in orbits that would have been detectable in the absence of stellar variabilities: thus either they have been swamped in (or confused with) the stellar variabilities and await confirmation with Doppler observations, or our model overestimates the number of BH systems on close-in orbits. In the latter case, the detection with \tess\ still remains to be plausible even if the actual rate is 10 times smaller than our estimate; if the overestimation factor is 100 or larger, the detection with the \tess\ data may not be feasible.


\vspace{0.5cm}
\section{Connections to X-ray Binaries}\label{sec:xb}

Given the lack of observational knowledge on the population of detached BH--star binaries, in Section \ref{sec:yields} we adopted simple models for the population to estimate the expected yields. On the other hand, we do have observational constraints on {\it interacting} systems, which are observed as X-ray binaries containing BHs (BHXBs). These BHXBs may be considered as the shorter-period end of the spectrum of detached systems, 
which can be extrapolated to give a rough estimate on the occurrence of tight but detached systems as discussed in this paper. Here we try this more empirical approach as an independent check of the feasibility. This exercise also reveals the potential of \tess\ to characterize X-ray binaries via optical light curves, as has routinely been performed with the ground-based photometry \citep[e.g.,][]{1978ARA&A..16..241B}. 

To do this, we count the number of BHXBs where the self-lensing and phase-curve signals induced on the stellar companion would be detectable with \tess, assuming that the system is in a quiescent phase and the system luminosity is not dominated by the accretion disk (i.e., we artificially ``turn off" interactions in these systems). We also take into account the incompleteness of the known BHXB population; those with weak X-ray emissions are likely more relevant for our estimate, but they are also more easily missed in the search for X-ray binaries.

\subsection{The Sample of X-ray Binaries with Confirmed BHs}

Our sample consists of $18$ X-ray binaries with dynamically confirmed BH companions in Table 1 of \citet{2006ARA&A..44...49R}, with two systems in the Large Magellanic Cloud being excluded. Distances to the systems are adopted from Table 4.1 of \citet{2006csxs.book..157M} except for GS 1354-64 \citep{2004ApJ...613L.133C}\footnote{\citet{2018arXiv180411349G} derived a much smaller distance than adopted in this work using the parallax from Data Release 2 of \gaia. The source of potential discrepancy is unclear, but here we simply adopt the previously estimated (larger) distance to give a conservative estimate.} and XTE J1650-500 \citep{2004ApJ...616..376O}. 

\subsection{Maximum Searchable Distance $d_{\rm max}$}

We adopt $P=0.8\,\mathrm{days}$ and $\mbh=7\,M_\odot$ as the representative values of the above BHXB sample, and assume a Sun-like star with $\ms=1\,M_\odot$ and $\rs=1\,R_\odot$ for a typical companion. For these parameters, the amplitudes of the self-lensing and phase-curve signals are $\approx 9\times 10^{-4}$ and $\approx 6\times 10^{-3}$, respectively, where the detectability of the latter is limited by the beaming signal (cf. Figure \ref{fig:amp_period}). Assuming that these signals are detectable when the signal-to-noise ratio of the phase-folded signal is larger than 10 for the white noise model in Section \ref{ssec:white_noise},
these amplitudes are roughly translated into the limiting magnitudes of $V\approx 11$ and $V\approx 15$, respectively, or the maximum searchable distances of $d_{\rm max}\approx 0.25\,\mathrm{kpc}$ and $d_{\rm max}\approx 1.3\,\mathrm{kpc}$ for the assumed Sun-like companion. These are shown with vertical bands in Figure \ref{fig:yield_bhxb}. Note that $d_{\rm max}$ for self-lensing is rather sensitive to the property of the stellar companion because the signal scales as $\rs^{-2}$ (Eqn. \ref{eq:sl}). It varies from $\approx0.1\,\mathrm{kpc}$ to $\approx0.4\,\mathrm{kpc}$ for A to K dwarf companions. The dependence is much smaller for the beaming signal (see Eqn.~\ref{eq:beam}).

\subsection{Cumulative Distribution of Distances to BHXBs}

Given the cumulative distance distribution of BHXBs, $N(d)$, the value of $N(d_{\rm max})$ roughly corresponds to the number of such systems whose self-lensing (if present) and phase-curve signals would be detectable with \tess, assuming no other source of optical light variations. We estimate $N(d)$ as $N(d)=fN_{\rm confirmed}(d)$, where $N_{\rm confirmed}(d)$ is the cumulative distance distribution of the above confirmed BHXB sample and $f$ is a correction factor for the incompleteness of the X-ray binary search. \citet{2018MNRAS.474...69A} evaluated the detectability of X-ray and optical signals as well as X-ray outbursts for the above BHXB sample, and concluded that the completeness of the detection is $\approx 1/30$ on average. We show $N(d)$ with $f=30$ motivated by this result as a thin histogram in Figure \ref{fig:yield_bhxb}, along with the filled histogram for $N_{\rm confirmed}(d)$.
For $d\lesssim1\,\mathrm{kpc}$ at which no BHXB has been detected, we extrapolate the distribution assuming $N(d)\propto d^2$, i.e., they are dominated by the disk population and their space density is roughly constant. Although the validity of these assumptions is uncertain, this extrapolation only matters the estimate regarding the self-lensing population.

\subsection{Results}

\begin{figure*}[ht!]
	\epsscale{0.8}
	\plotone{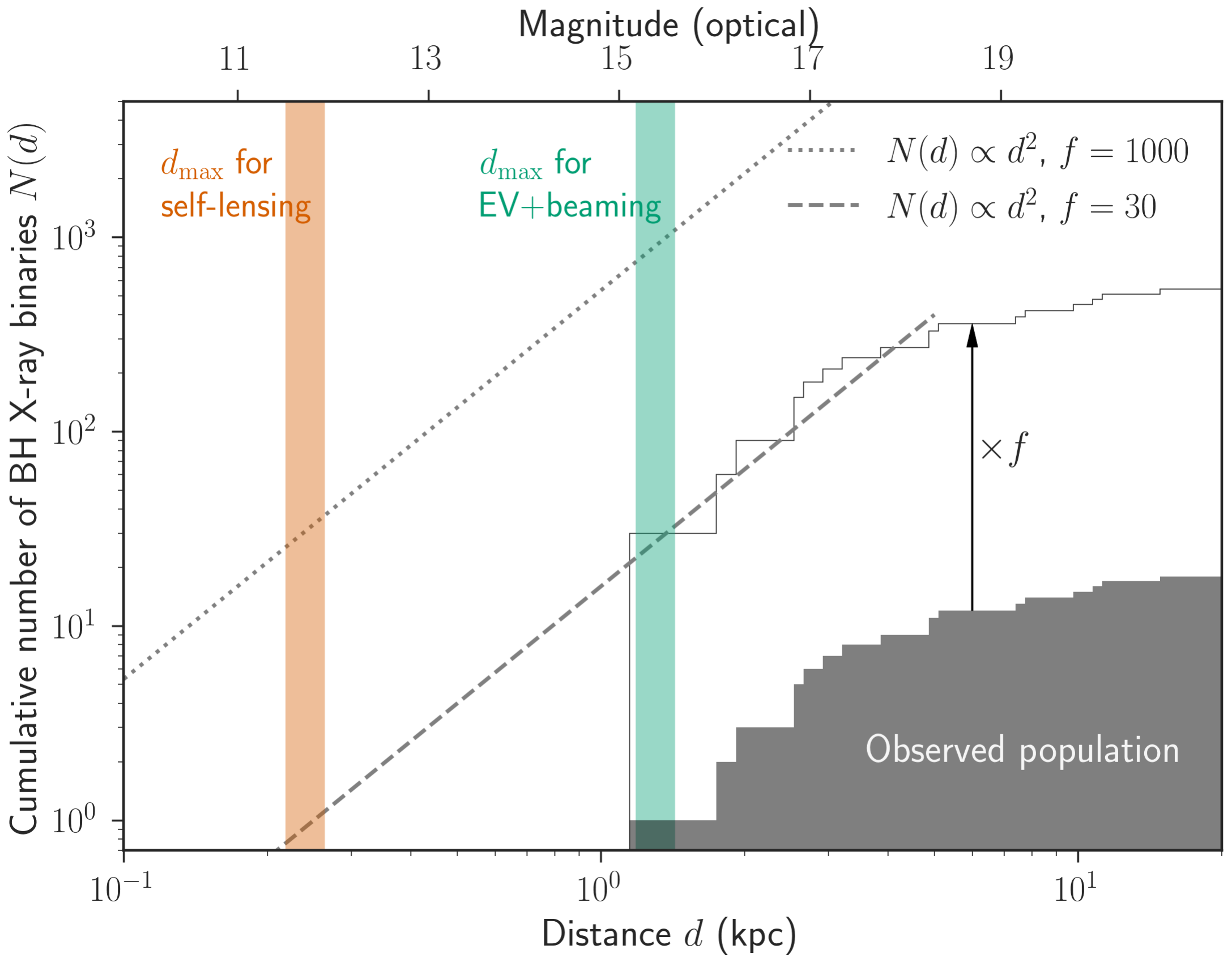}
	\caption{Estimates for the cumulative distribution of distance $d$ to X-ray binaries with BHs, $N(d)$. Maximum searchable distances $d_{\rm max}$ for the self-lensing and phase-curve signals are also shown for $\mbh=7\,\ms$, $P=0.8\,\days$, and $\ms=1\,\msun$ (vertical bands). The value of $N(d_{\rm max})$ corresponds to the number of systems whose optical signals would be detectable with \tess\ assuming no other source of optical signal. The gray histogram in the bottom right is for the observed BHXB sample, and thin histogram is the distribution inflated by a factor of $f$ to take into account the incompleteness of the X-ray binary detection. The dashed and dotted lines show the extrapolation of this distribution for $f=30$ and $f=1000$, respectively, assuming $N(d)\propto d^2$.\label{fig:yield_bhxb}
	}
\end{figure*}

Figure \ref{fig:yield_bhxb} shows that the phase-curve signal will be detectable among several tens of BHXB systems for $f=30$. While we have focused only on BHXBs, the number of potentially accessible systems doubles if we also consider X-ray binaries with confirmed NSs or candidate BHs \citep{2006csxs.book..157M, 2017hsn..book.1499C}. Moreover, there are a few hundreds of sources whose detailed properties are still unclear \citep{2006A&A...455.1165L, 2007A&A...469..807L}. Thus we expect that \tess\ photometry will be useful for characterizing at least some X-ray binaries with phase-curve variations. For the self-lensing signal, on the other hand, $N(d_{\rm max})$ is $\mathcal{O}(1)$ or smaller. This number needs to be down-weighted by the eclipse probability, which is typically $\mathcal{O}(0.1)$, to estimate the number of actual detections. Thus the detection of self-lensing signals in X-ray binary like systems is not promising for $f=30$. 


If we set $f=1000$ (dotted line in Figure \ref{fig:yield_bhxb}), we expect a few hundred systems showing detectable phase-curve signals and a few with self-lensing signals (after taking into account the eclipse probability). These numbers are roughly consistent with the estimates in the corresponding ($P$, $\mbh$) range from our population models in Section \ref{sec:yields} (Figures \ref{fig:yield} or \ref{fig:yield_ce}). This suggests that our population model is roughly consistent with the observed BHXB population if there exist $\approx 30$ times more detached BH--star systems than X-ray systems. The value appears to be compatible with the result of a population synthesis study \citep{2002ARep...46..667T} in the order of magnitude sense. 

\vspace{0.5cm}
\section{Summary and Outlook}\label{sec:summary}

We estimated that the self-lensing and phase-curve signals induced by BH companions on tight ($\lesssim0.3\,\mathrm{au}$) but detached orbits will be strong enough to be detectable in the \tess\ light curves of effectively $\sim 10^5$ and $\sim 10^6$ stars, respectively (Figure \ref{fig:ns}), taking into account inclination dependence of the signals. The detectability of these signals were evaluated using the stellar properties in the \tess\ input catalog \citep{2017arXiv170600495S}, \tess\ noise model in \citet{2015ApJ...809...77S}, and the light curves of spotted stars from the prime \kepler\ mission \citep{2014ApJS..211...24M} to gauge the impact of stellar activities.
If combined with simple models for the population of detached BH--star binaries (Figure \ref{fig:bhpops}), these ``searchable" stars are expected to host $\sim 10$ and $\sim100$ detectable BH companions (Figures \ref{fig:yield} and \ref{fig:yield_ce}), although we cannot exclude the possibility that the latter may be associated with a comparable number of false-positives due to stellar activities. The most promising targets turned out to be BHs with masses of a few $10\,\msun$ and orbital periods of a few days. This large population of BHs, if identified, will reveal the BH mass function down to $\sim1\,\msun$ range, semi-major axis/eccentricity distribution of their orbits, positions and velocities of the systems in the Galaxy, and chemical compositions of the companion stars. These constraints will be valuable to probe mass ejection and natal kick during the BH formation, as well as the binary evolution process that may result in the observed close-in orbits. Since non-zero detections are expected from our models, even the null detection, if quantified, will provide critical information on the models of interacting binaries containing BHs. Although we have focused on systems with BHs, the \tess\ light curves are also sensitive to NS companions down to $\sim1\,\msun$ (Figure \ref{fig:ns}). We may also detect phase-curve signals from optical counterparts of X-ray binaries in a quiescent phase, which potentially allow for better characterization of both known and unknown X-ray binaries.

Suppose that candidate BH companions are identified from the \tess\ photometry, what needs to be done next? The self-lensing signal, if identified, will provide the least ambiguous targets that are also best suited for further characterization: the precise orbital inclination and period from the light curve allow for the precise mass determination with radial velocity measurements. The stellar radius estimate using the \gaia\ parallax, if available, even allows for mass determination with light curves alone; because the pulse height constrains $\mbh/\rs^2$, the BH mass can be determined at least to the precision of $20\%$, or even better if spectroscopic effective temperature of the companion is available. 
The candidates identified with the phase-curve signals would require further vetting with follow-up spectroscopy to confirm their ``SB1" nature and to measure spectroscopic orbits.  While the candidates identified from either method will most likely be bright enough for follow-up spectroscopy, the archival data from large spectroscopic surveys, such as APOGEE \citep{2018ApJS..235...42A}, RAVE \citep{2017AJ....153...75K}, LAMOST \citep{2012RAA....12.1197C}, and GALAH \citep{2018MNRAS.478.4513B}, will also play an essential role to complement the \tess\ photometry, both in terms of target vetting and dynamical/chemical characterization. Indeed, those archival data alone might even provide sufficient information to confirm or reject some of the candidates. Eventually, \gaia\ will also provide further information based on astrometric orbits and/or multi-epoch radial velocity measurements.


There are also other classes of objects that can be searched with similar methods. This includes BH/NS companions of WDs, which were not considered in this paper because the eclipse probability is small, timescales of the detectable signals may be too short for the $30$-minute cadence photometry, and most of them are likely too faint for \tess. Nevertheless, they may still be good targets for all-sky photometric surveys from the ground \citep[cf.][]{2002A&A...394..489B}. If the \tess\ mission is extended, and/or for stars in the overlapping regions of observing sectors, BH/NS companions on longer-period orbits around evolved stars, as identified in \citet{2018arXiv180602751T} and \citet{2019ApJ...872L..20G}, will also be within reach.

\acknowledgments

We thank Josh Winn, Adrian Price-Whelan, and Hajime Kawahara for helpful conversations. We thank Almog Yalinewich and Tsevi Mazeh for comments on the manuscript. 
This paper includes data collected by the \kepler\ mission. Funding for the \kepler\ mission is provided by the NASA Science Mission directorate. 
Work by K.M. was performed under contract with the California Institute of Technology (Caltech)/Jet Propulsion Laboratory (JPL) funded by NASA through the Sagan Fellowship Program executed by the NASA Exoplanet Science Institute. 





\bibliographystyle{aasjournal}


\listofchanges

\end{document}